\documentclass[a4,twocolumn]{article}
\usepackage{natbib}
\usepackage{graphicx}
\usepackage{color} 
\usepackage{verbatim,framed}
\usepackage{url}

\definecolor{aqgr} {rgb}{0.0,1.0,0.6} 
\definecolor{viol} {rgb}{0.8,0.6,1.0}
\definecolor{coldr}{rgb}{1.0,1.0,1.0} 
\definecolor{colwp}{rgb}{1.0,1.0,1.0} 
\definecolor{colok}{rgb}{1.0,1.0,1.0} 

\title{A hypothesis on the  \\
       role of transposons}

\author{Alessandro Fontana$^{1}$ \\
\mbox{}\\
$^1$IEEE \\
alessandro.fontana@ieee.org}

\begin{document}
\maketitle

\begin{abstract}
``Epigenetic Tracking'' is an evo-devo method to generate arbitrary 2d or 3d shapes; as such, it belongs to the field of ``artificial embryology''. {\em In silico} experiments have proved the effectiveness of the method in devo-evolving shapes of any kind and complexity (in terms of number of cells, number of colours, etc.), establishing its potential to generate the complexity typical of biological systems. Furthermore, it has also been shown how the underlying model of development is able to produce the artificial version of key biological phenomena such as embryogenesis, the presence of junk DNA, the phenomenon of ageing and the process of carcinogenesis. In this paper the evo-devo core of the method is explored and the result is a novel hypothesis on the biological role of genomic transposable elements, according to which transposition in somatic cells during development drives cellular differentiation and transposition in germ cells is an indispensable tool to boost evolution. Thus transposable elements, far from being ``junk'', have one of the most important roles in multicellular biology.
\end{abstract}

\section{Introduction}

\colorbox{colok}{The previous work} in the field of Artificial Embryology (see \citep{AY03SM} for a comprehensive review) can be divided into two broad categories: the grammatical approach and the cell chemistry approach. In the grammatical approach development is guided by sets of grammatical rewrite rules; context-free or context-sensitive grammars, instruction trees or directed graphs can be used; L-systems were first introduced by Lindenmayer \citep{AY68LX} to describe the complex fractal patterns observed in the structure of trees. The cell chemistry approach draws inspiration from the early work of Turing \citep{AY52TX}, who introduced reaction and diffusion equations to explain the striped patterns observed in nature (e.g. shells and animals' fur); this approach attempts to simulate cell biology at a deeper level, going inside cells and reconstructing the dynamics of chemical reactions and the networks of chemical signals exchanged between cells.

\colorbox{colwp}{``Epigenetic Tracking'' is} the name of an embryogeny applied to morphogenesis, i.e. the task of generating arbitrary 2d or 3d shapes, described in \citep{AY08AX}. From this initial work, two lines of research are possible. One tries to make use of the method as a general-puropose tool to solve real-world problems; another line of research tries to bridge the gap between the model and real biology. This second research direction has been pursued in \citep{AY09AX} and \citep{AY10A1}, where the model has been shown able to provide insights into key aspects of biology -such as the phenomenon of ageing and the process of carcinogenesis- and will be continued in this paper, which deals with another feature of biological systems: the ubiquitous presence in the genomes of most species of pieces of DNA called ``transposable elements'', capable of moving between different chromosomal loci. The rest of this paper is organised as follows: section 2 describes concisely the model of development; sections 3 describes a procedure called ``Germline Penetration'', which lies at the heart of the evo-devo process; section 4 analyses analogies and differences between driver cells and stem cells; section 5 outlines the biological role of transposons; section 6 draws the conclusions. 

\section{Epigenetic Tracking highlights}

\begin{figure}[t!] \begin{center}
{\fboxrule=0.2mm\fboxsep=0mm\fbox{\includegraphics[width=8.60cm]{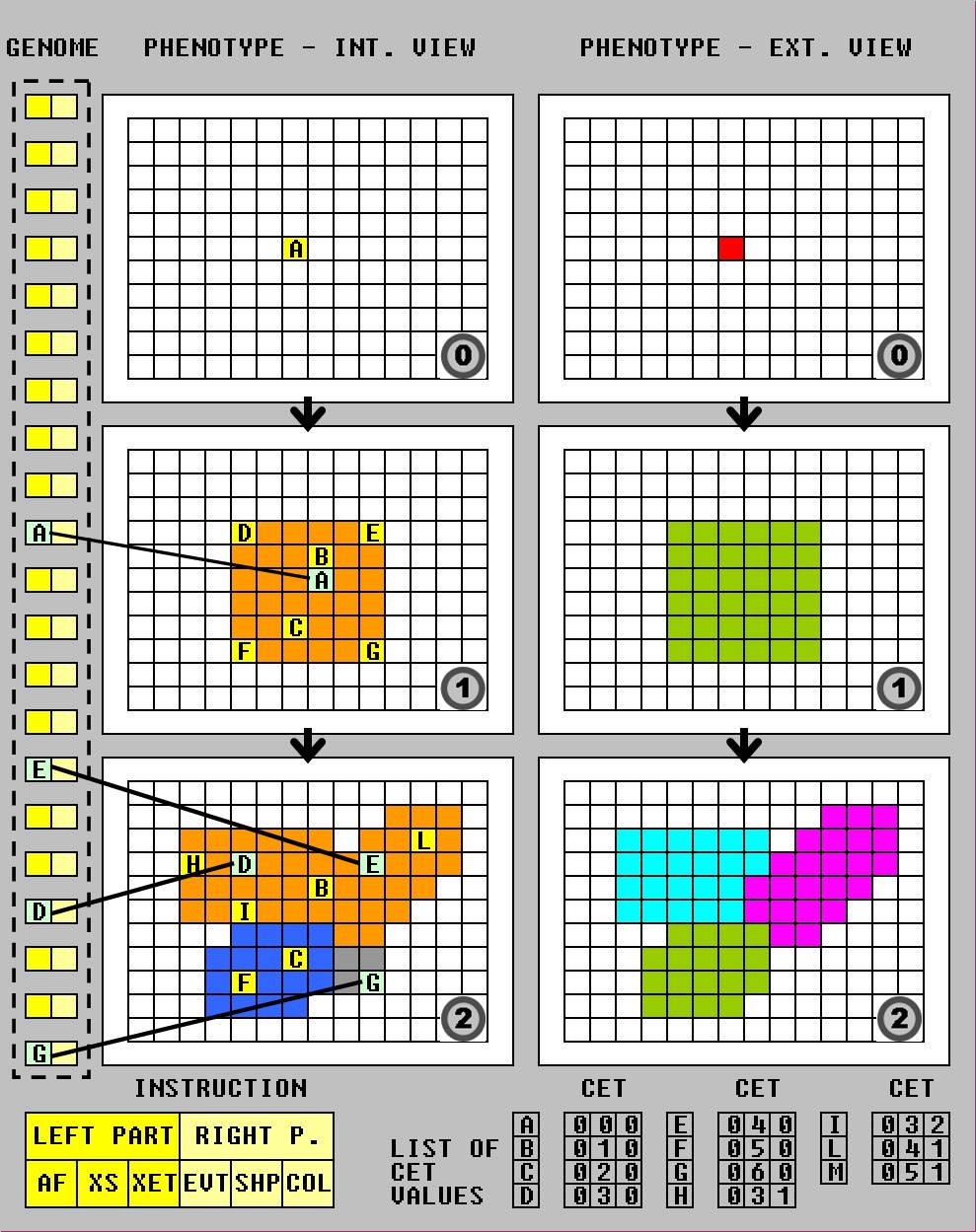}}}
\caption{Example of development in three steps (AS=0,1,2) driven by three instructions: a proliferation triggered in step 1 on the driver cell labelled with A, an apoptosis triggered in step 2 on the driver cell labelled with D and another proliferation triggered in step 2 on the driver cell labelled with E. Internal view on the left, external view on the right.}
\label{figxx}
\end{center} \end{figure}

\colorbox{colok}{Shapes are composed} of cells deployed on a grid; development starts with a cell (zygote) placed in the middle of the grid and unfolds in N age steps, counted by the variable ``Age Step'' (AS), which is shared by all cells and can be considered the ``global clock'' of the organism. Cells belong to two distinct categories: ``normal'' cells, which make up the bulk of the shape and ``driver'' cells, which are much fewer in number (typical value is one driver each 100 normal cells) and are evenly distributed in the shape volume. Driver cells have a Genome (an array of ``instructions'', composed of a left part and a right part) and a variable called cellular epigenetic type (CET, an array of integers). While the Genome is identical for all driver cells, the CET value is different in each driver cell; in this way, it can be used by different driver cells as a ``key'' to activate different instructions in the Genome. The CET value represents the source of differentiation during development, allowing driver cells to behave differently despite sharing the same Genome. A shape can be ``viewed'' in two ways: in ``external view'' cells are shown with their colours; in ``internal view'' colours represent cell properties: blue is used for normal cells alive, orange for normal cells just (i.e. in the current age step) created, grey for cells that have just died, yellow for driver cells (regardless of when they have been created).

\colorbox{colok}{An instruction's left} part is composed of the following elements: an activation flag (AF), indicating whether the instruction is active or not; a variable called XET, of the same type as CET; a variable called XS, of the same type as AS. At each step, for each instruction and for each driver cell, the algorithm tests if the instruction's XET matches the driver's CET and if the instruction's XS matches AS. In practise, XS behaves like a timer, which makes the instruction activation wait until the clock reaches a certain value. If a match occurs, it triggers the execution of the instruction's right part, which codes for three things: event type, shape and colour. Instructions give rise to two 'types' of events: ``proliferation instructions'' cause the matching driver cell (called ``mother cell'') to proliferate in the volume around it (called ``change volume''), ``apoptosis instructions'' cause cells in the change volume to be deleted from the grid; the parameter 'shape' specifies the shape of the change volume, in which the proliferation/apoptosis events occur, choosing from a number of basic shapes called ``shapint primitives''; in case of proliferation, the parameter 'colour' specifies the colour of the new cells.

\begin{figure}[t!] \begin{center}
{\fboxrule=0.2mm\fboxsep=0mm\fbox{\includegraphics[height=8.00cm]{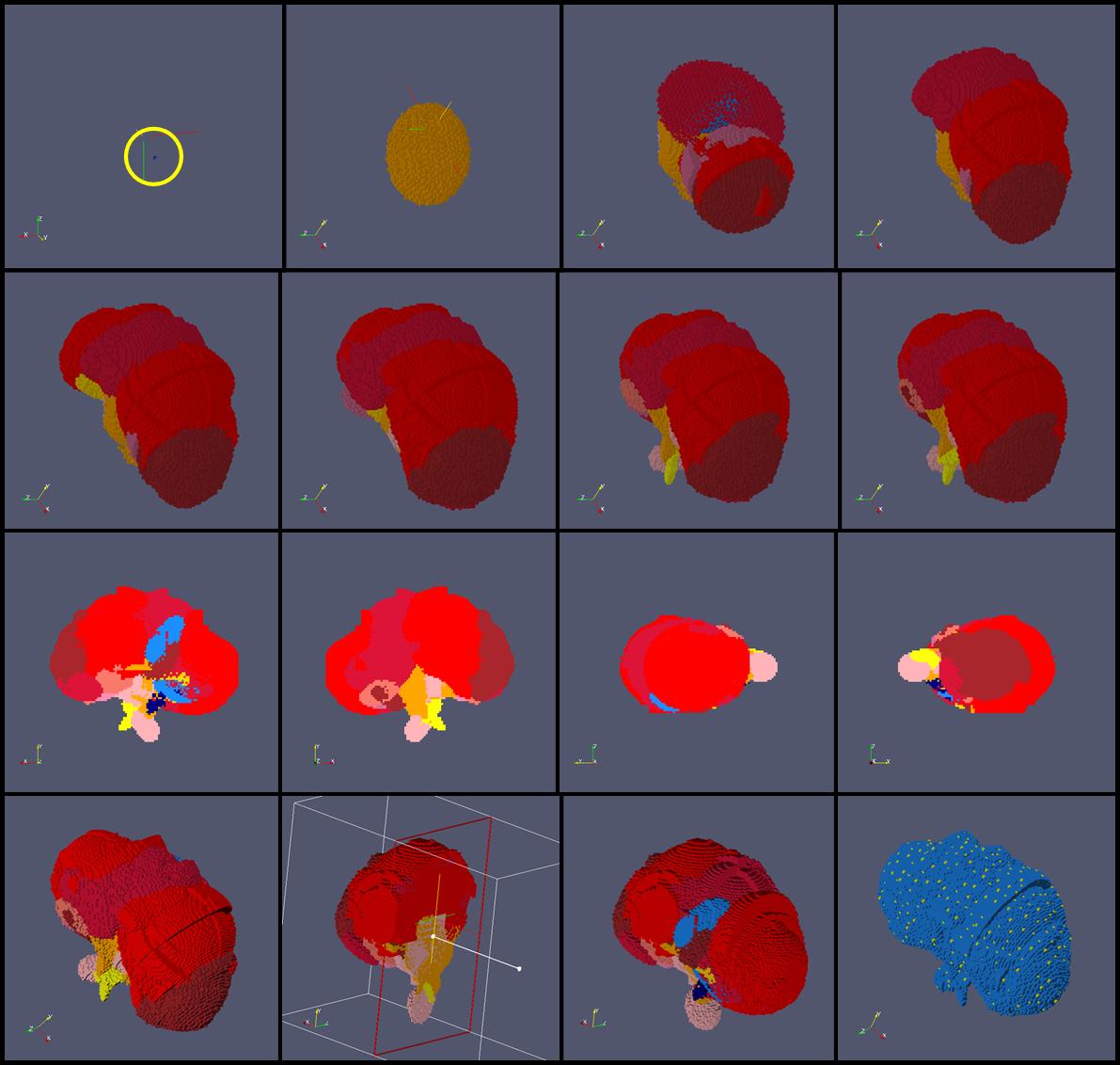}}}
\caption{Example of development coded in a Genome composed of 360 instructions, evolved in 30000 generations; the shape represents a kidney, composed of 150000 cells. In the upper part, the development sequence; in the lower part, some snaphots of the final phenotype taken from different angles.}
\label{figxx}
\end{center} \end{figure}

\colorbox{colok}{Always in case} of proliferation, both normal cells and driver cells are created: normal cells fill the change volume, driver cells are ``sprinkled'' uniformly in the change volume. To each new driver cell a new, previously unseen and unique CET value is assigned (consider for example proliferation triggered on A in figure 1), obtained starting from the mother's CET value (the array [0,0,0] in the figure, labelled with A) and adding 1 to the value held in the ith array position at each new assignment (i is the current value of the AS counter); with reference to the figure, the new driver cells are assigned the values [0,1,0],[0,2,0],[0,3,0], ... , labelled with 'B','C','D', etc. (please note that labels are just used in the figures for visualisation purposes, but all operations are made on the underlying arrays). In practise a proliferation event does two things: first creates new normal cells and sends them down a differentiation path (represented by the colour); then creates other driver cells, one of which can become the centre of another event of proliferation or apoptosis, if in the Genome an instruction appears, whose XET matches such value. Figure 1 reports an example of development hand-coded. 

\colorbox{colok}{The model of} development described, coupled with a standard Genetic Algorithm (GA), becomes an evo-devo method to generate arbitrarily shaped 2d or 3d cellular sets. The method evolves a population of Genomes that guide the development of the shape starting from a zygote initially present on the grid, for a number of generations; at each generation development is let unfold for each Genome and, at the end of it, adherence of the shape to the target shape is employed as fitness measure. {\em In silico} experiments (an example in figure 2) have proved the effectiveness of the method in devo-evolving any kind of shape, of any complexity (in terms e.g. of number of cells, number of colours, etc.); being shape complexity a metaphor for organismal complexity, such simulations established the method's potential to generate the complexity typical of biological systems. The power of the method depends essentially on the features of the model of development, but the speed of the evolutionary process is also safeguarded by a special procedure, which constitutes the subject of next section.  

\section{Germline Penetration and the evo-devo ``core''}

\colorbox{colwp}{In order to} develop a given shape, the Genetic Algorithm has to come up with instructions whose XET value matches a CET value belonging to one of the drivers present in the shape volume. With realistic values for the XET array size (greater than 2-3), the space the GA has to search becomes large enough to bring the evolutionary process to a halt (if the array is made up of 10 scalars and each scalar can assume 10 possible values, the search space size is $10^{10}$). A possible countermeasure would be to ``suggest'' to the GA XET values that are guaranteed or very likely to match existing CET values (instead of having it guessing them). If we suggest to the GA to try as XET values elements drawn from the set of CET values generated during development, the match is guaranteed. This idea is implemented in a procedure called ``Germline Penetration'', executed at the end of each individual's development, which copies at random (some) CET values generated during the individual's development onto XET values of instructions in a copy of the Genome called ``germline'' Genome, destined to become the (``somatic'') Genome of next generation.

\colorbox{colwp}{An example of} how Germline Penetration works is provided in figure 3, which shows the development sequence for a shape belonging to ``species'' X, at generation K. For this species development consists of a single change event triggered on the zygote and occurring in step 1, in which a number of new CET values are generated: B, C, D, etc.; to further develop the shape, some new instructions must be cast on these new CET values. As we said, guessing them is an almost impossible undertaking for the GA: Germline Penetration intervenes copying some of the new CET values into XET values of instructions present in the germline Genome, where they find their way into the Genome of next generation's individuals. Evolution, whose objective is to produce individuals with a high fitness level, is now provided with ``good'' instructions' left parts: it has now to optimise the relevant right parts, a process that can take several generations.

\colorbox{colwp}{This is what} happens after H generations (at generation K+H) for the individual shown in figure 4: two instructions are now triggered on CET values D and E (which had been transferred into the Genome by Germline Penetration), giving origin to as many change events occuring in step 2: the individual belongs to a new species, Y. The new CET values generated as a result of the new events occurred in step 2 are again migrated to the germline Genome to be embedded into the Genome of the offspring and the whole cycle repeats itself. The instructions with the copied XET values are initially set as inactive (parameter AF is set to 'OFF'), otherwise they would all become active with a non-optimised right part and development would be disrupted: their activation, obtained through a ``flip'' of the activation flag AF, is left to a subsequent genomic mutation. As a result, at any given time in the course of evolution, most instructions in Genome are inactive and, in analogy with real genomes, can be defined ``junk'' instructions. 

\begin{figure}[t!] \begin{center}
{\fboxrule=0.2mm\fboxsep=0mm\fbox{\includegraphics[width=8.60cm]{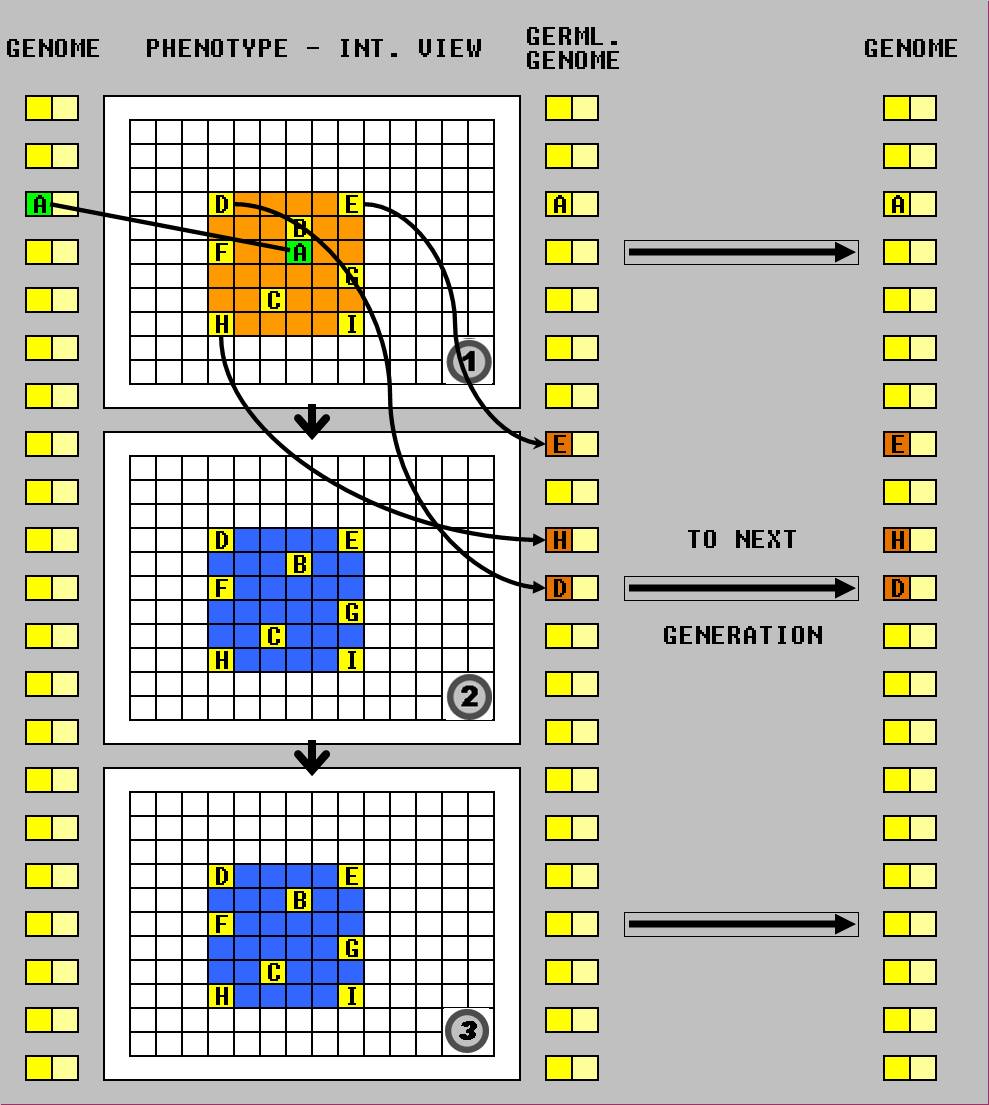}}}
\caption{Germline Penetration in action: development of a species X individual, generation K. Development stops at step 1; some of the CET values generated during development leave the respective driver cells and are conveyed towards the germline Genome and copied onto instructions' XET values. The germline Genome is passed on, to become the Genome of next generation individuals: this Genome -incorporated in all cells- contains the new XET values.}
\label{figxx}
\end{center} \end{figure}

\begin{figure}[t!] \begin{center}
{\fboxrule=0.2mm\fboxsep=0mm\fbox{\includegraphics[width=8.60cm]{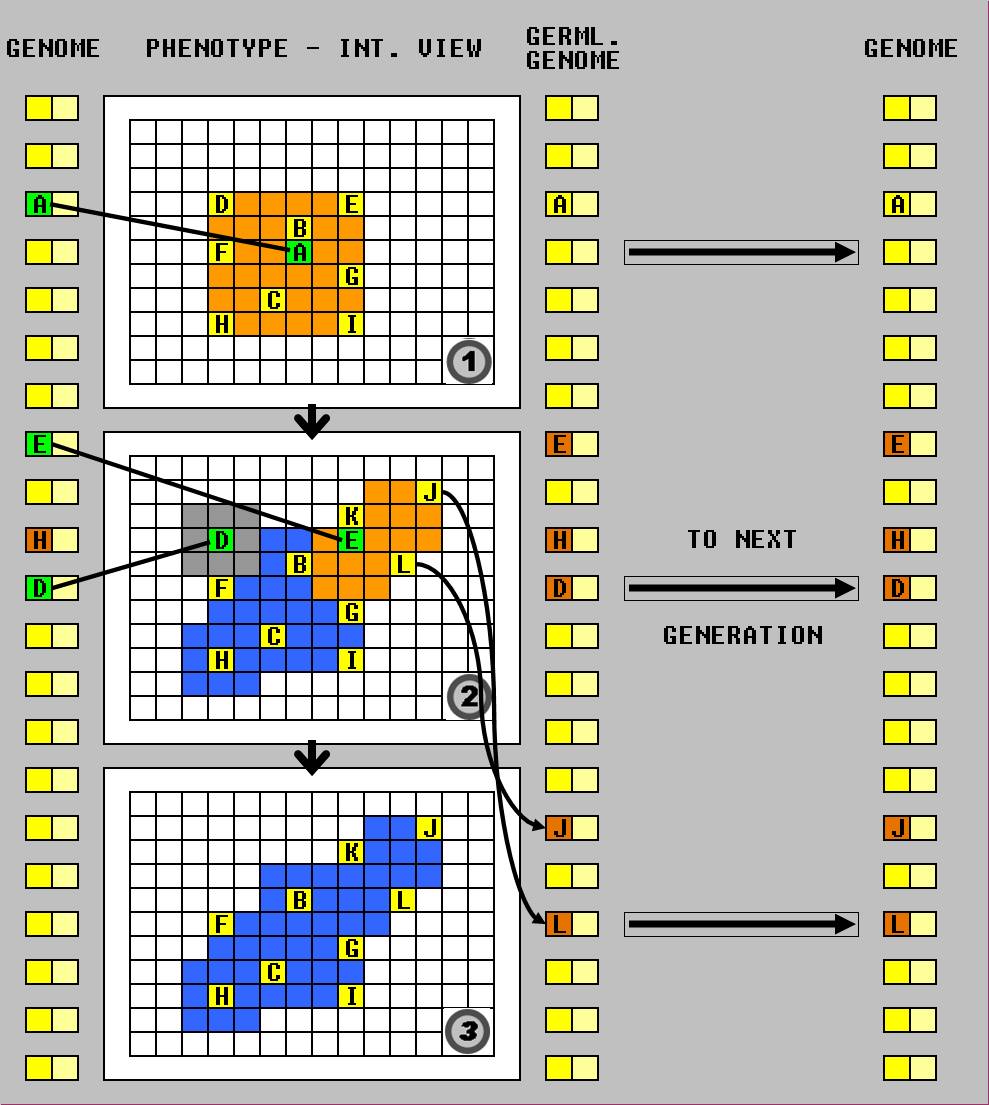}}}
\caption{Germline Penetration in action: development of a species Y individual, generation K+H. New instructions, derived from the transplanted elements, are present in the Genome, whose XET values match some of the CET values generated in step 1; such instructions carry on the development of species X, giving rise to a new species (Y). The new CET values generated in step 2 are again copied into the germline Genome and passed to the next generation.}
\label{figxx}
\end{center} \end{figure}

\colorbox{colwp}{Figures 3 and} 4 provide a snapshot of the evo-devo core of Epigenetic Tracking: a driver cell induces a proliferation, committing the cells created to specific fates; as a result, some new driver cells/CET values are generated; the CET values find their way into the germline Genome where they become incorporated in instruction's left parts, to be passed on to the next generation's individuals; after a number of generation evolution finds suitable solutions for the relevant right parts and development can move ahead. In this view development and evolution appear to be two inextricably intertwined sides of the same process (linked together by Germline Penetration), not unlike electricity and magnetism are two manifestations of the same physical phenomenon, governed by a common set of laws: the sentence ``nothing in biology makes sense except in the light of evolution'' could be rephrased as ``nothing in multicellular biology makes sense except in the light of devo-evolution''.

\colorbox{colwp}{The Germline Penetration} procedure, introduced as a means to speed-up evolution, achieves the objective transplanting into the Genome regulatory elements corresponding to driver cells present in the organism. We know that, whenever a driver cell is triggered to proliferate by means of a suitable instruction, a ``wave'' of new CET values is created in the body of the (new) species; the action of Germline Penetration translates this wave into a corresponding wave of new XET values spreading in the Genome. The moments of such occurrences during evolution coincide with milestones in which radical changes occur to the species being evolved, causing new body parts, or new features to appear. In other words, the spreading in the Genome of new waves of regulatory elements in the course of evolution corresponds to moments in which new branches (representing new species) are generated in the artificial ``tree of life''. After this journey in the artificial world, it is now time to come back to earth and see what lessons can be learned for real biological systems.    

\section{Embryogenesis and stem cells}

\colorbox{colwp}{The model of} cellular growth called ``Epigenetic Tracking'' has been tested experimentally with the problem of artificial morphogenesis and differentiation, implemented respectively through the shaping and colouring of cellular sets: therefore its interpretation as a model of morphogenesis and cellular differentiation is straightforward. A key element in the cellular model described is represented by driver cells, a subpopulation of cells which ``drive'' development. Only driver cells can be instructed to develop (proliferate or undergo apoptosis) by the Genome: they represent the scaffolding, the backbone of the developing shape and make it possible to steer development by acting on a small subset of cells. The CET value stored inside the cell (and moved along with the cell) takes different values in different driver cells and represents the source of differentiation during development. This feature represents a key difference with respect to other cellular models that rely on positional information and chemical micro-environment to provide the information necessary for differentiation.

\colorbox{colwp}{Stem cells are} found in most multi-cellular organisms; the classical definition of a stem cell requires two properties: i) {\bf self-renewal} and ii) {\bf potency}. Two types of stem cells exist: embryonic stem cells (found in the inner cell mass of the blastocyst) and adult stem cells (found in adult tissues). {\bf Embryonic stem (ES) cells} are totipotent: this means they are able to differentiate into all cell types of the body; in order to maintain the undifferentiated state, ES cells must be kept under tightly regulated culture conditions, otherwise they rapidly differentiate. {\bf Adult stem cells} are pluripotent undifferentiated cells found throughout the body after embryonic development that divide, to replenish dying cells and regenerate damaged tissues; pluripotency distinguishes adult stem cells from totipotent embryonic stem cells: they can only form a limited set of cell types. 

\colorbox{colwp}{Since driver cells} guide development, out of which all artificial cell types are generated, it is quite natural to think of driver cells as the artificial equivalent of embryonic stem cells. There are also many analogies between the concept of driver cell and the concept of Spemann's organiser: for instance, if a driver cell (a Spemann's organiser) destined to give rise to a certain shape (embryo) part is moved to a different position of the growing shape (embryo), that shape (embryo) part will grow in the new, ectopic position. As stem cells, also driver cells are characterised by a decreasing degree of potency as development progresses; as far as the self-renewal property is concerned, the situation is a bit more complex: in fact a driver cell, when proliferating, generates other driver cells, whose CET values however are not the same as the mother's. At present adult stem cells have no correspondent in the model; this gap can nevertheless be easily filled introducing a new type of driver cell that, when proliferating, generates among the others also driver cells having the CET value of the mother: we can call such driver cells ``maintenance'' driver cells, while standard driver cells will be called ``development'' driver cells (see figure 5).

\colorbox{colwp}{As a result of} these considerations, in place of the traditional distinction between embryonic stem cells, adult stem cells and all other cells, which is based on time, we are led to propose a new classification of biological cells, based on function: i) ``development'' stem cells are a subset of cells which drive development, present throughout the entire organism's life (in the embryo earliest stages they coincide with embryonic stems cells); ii) ``maintenance'' stem cells are a subset of cells which divide to regenerate damaged tissues, present from a given development stage onwards, for the rest of the organism's life (in the adult they coincide with adult stems cells); iii) other, ``normal'' cells, characterised by a certain degree of ``plasticity'', i.e. the susceptibility of being turned (by stem cells) into a number of cell types. In other words, development stem cells correspond to development driver cells and maintenance stem cells correspond to maintenance driver cells. The key element emerging from this re-classification is that, unlike embryonic stem cells which are present in the embryo's earliest stages and then seem to ``dissolve'' as development progresses, development stem cells are present for the whole duration of the organism life, from the zygote stage to the moment of death, a property that has been shown to have profound implications \citep{AY09AX} \citep{AY10A1}.

\begin{figure}[t!] \begin{center}
{\fboxrule=0.2mm\fboxsep=0mm\fbox{\includegraphics[width=8.00cm]{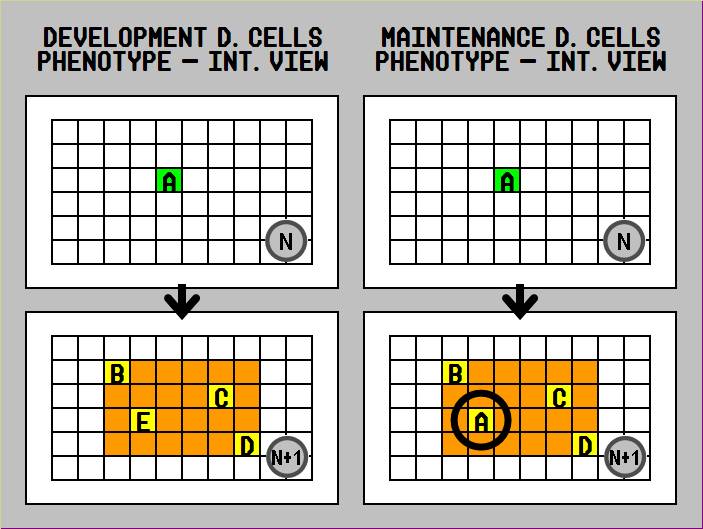}}}
\caption{``Development'' driver cells (on the left) and ``maintenance'' driver cells (on the right). In the progeny of the latter one or more drivers with the same CET of the mother are present.}
\label{figxx}
\end{center} \end{figure}


\section{The biological role of transposons}

\colorbox{colwp}{Transposable elements (TE),} or transposons, first discovered by B. McClintock \citep{EX50MC}, are sequences of DNA that can move around to different positions in the genome, a process called transposition; in the process, they can cause mutations, chromosomal rearrangements and lead to an increase of genome size. Transposons can be categorised based on their mechanism of transposition. Class I transposons, or retrotransposons, copy themselves by first being transcribed to RNA, then reverse transcribed back to DNA (by reverse transcriptase), and then being inserted at another position in the genome; this mechanism of transposition can be described as ``copy and paste''; major subclasses of retrotransposons are represented by LTR retrotransposons, long interspersed nuclear elements (LINE's) and short interspersed nuclear elements (SINE's); the most common SINE's in primates are called ``Alu sequences''. Class II transposons, or DNA transposons, move directly from one genome position to another using a transposase, with a mechanism that can be characterised as ``cut and paste''. Transposons represent a large fraction of a genome (30-40\% in mammals): the amount of seemingly useless material initially led researchers to call it ``junk DNA'', until further research hinted that it could indeed have a biological role. Transposable elements have been studied along two main dimensions: development and evolution.

\colorbox{colwp}{A first line} of research has been concerned with the dynamics of TE diffusion in genomes across multiple generations, as can be evidenced through modern genome-wide analysis techniques; in this light transposable elements are mostly interpreted as genome parasites, in accordance with the ``junk DNA'' hypothesis. Transposons are associated to major evolutionary changes \citep{EX05FL} \citep{EX10OG}; many transposable elements are present only in specific lineages (Alu in primates for example), implying that the TE colonisation of the genome and the branching of the relevant lineage have occurred simultaneously; this remarkable correlation suggests indeed a causal link between the appearance of transposons in the genome and the evolutionary change that originated the lineage. A good example is provided by the hominoid lineage: recent findings indicate that periodic expansions of LINE's and SINE's correspond temporally with major divergence points in the hominoid evolutionary path. On the other hand, lineages whose genomes are not subject to intermittent TE infiltrations appear to be more static from an evolutionary point of view. A specific increased activity of some TE families in the germline has been observed \citep{EX07MM}, which is coeherent with the hypothesised evolutionary role of transposons: in fact, only when transposons become fixed in the germline, they can be passed on to the next generation and have trans-generational effects.

\colorbox{colwp}{Transposable elements have} have been initially characterised as elements controlling phenotypic characteristics during development in maize \citep{EX50MC}. In the course of plant development, they can insert themselves near genes regulating pigment production in specific cells and their descendant cell lineages, inhibiting their action and making the cells unable to produce the pigment: the overall macroscopic effect is an uneven pigment distribution in sectors of the plant (see figure 6). This regulatory role is supported by further research indicating that (in human) transposons tend to be located near genes known to play a role in development \citep{EX07LB}. Some TE families undergo changes in their methylation status during development: whereas Alu's, for instance, are almost completely methylated in somatic tissues, they are hypomethylated in the male germline and tissues which rely on the genome paternal half for development \citep{EX93HB}. Recent observations corroborate the view that TEs are active in somatic cells, opening the possibility that they can bring diversity among somatic cells having the same genome \citep{EX07CL}; specifically, a human L1 element was shown to retrotranspose in vitro in rodent neural progenitor cells (NPCs) and in vivo in mouse brain. All these elements concur to indicate that TE's can be grouped in two broad categories: i) TE's active in somatic cells during development and ii) TE's, active in germline cells across multiple generations. 

\begin{figure}[t!] \begin{center}
{\fboxrule=0.2mm\fboxsep=0mm\fbox{\includegraphics[width=8.40cm]{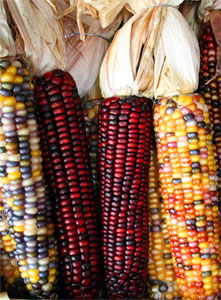}}}
\caption{The effects of transposition in maize. Transposons inserted near genes controlling pigment production can block or inhibit their activity. For example, if the transposon moves to a position adjacent to a pigment-producing gene, the cell is unable to produce the purple pigment. Transposons activated only in specific cells or sectors during development can produce macroscopic effects, such as a non uniform pigmentation of the corn.}
\label{figxx}
\end{center} \end{figure}

\colorbox{colwp}{In the biological} interpretation provided in the previous section, driver cells have been hypothesised to correspond to stem cells; a key ingredient of driver cells, the CET value, has nonetheless remained without a biological counterpart. The CET value is the ``code'' that distinguishes different driver cells; it matches with the XET value in the left part of a change instruction, from which a change event, such as proliferation or apoptosis, is produced; in case of proliferation, beside the bulk of normal cells which are being sent down a differentiation path, other driver cells are created and a corresponding number of CET values are assigned to them: these codes represent the ``handle'' by means of which such driver cells can be given other instructions to execute at a subsequent development step. From what we said, it is clear that the importance of the CET value cannot be overstated: its presence is an absolutely essential ingredient of Epigenetic Tracking, without which the method would not be able to function; we can make one further step and say that the role of the CET value is so crucial for artificial development, that it must necessarily be present also in natural development.  

\begin{figure}[t!] \begin{center}
{\fboxrule=0.2mm\fboxsep=0mm\fbox{\includegraphics[width=8.00cm]{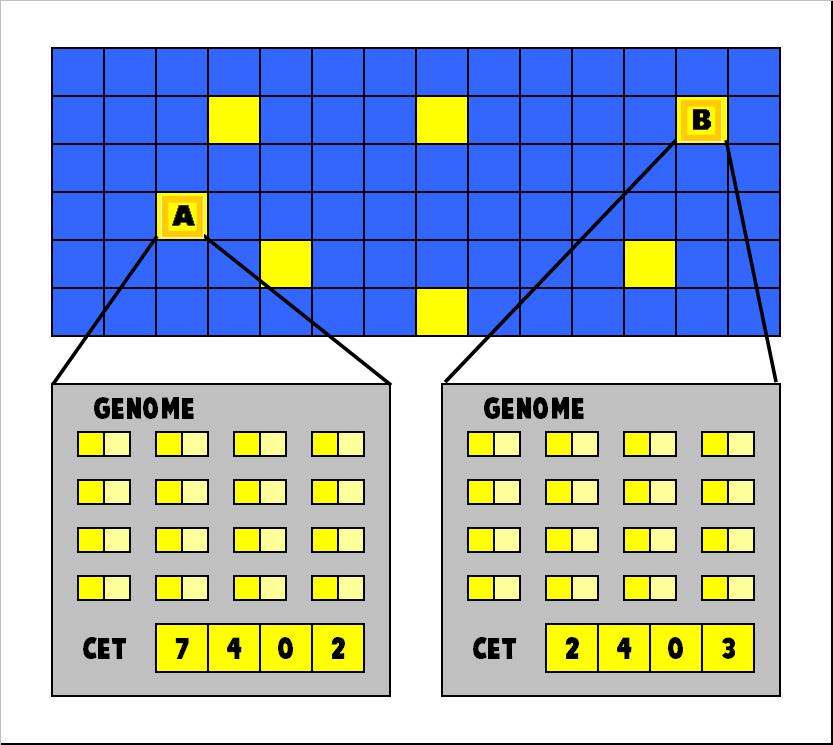}}}
\caption{Different CET values in two driver cells in Epigenetic Tracking; the CET value is stored in a contiguous portion of computer memory.}
\label{figxx}
\end{center} \end{figure}

\colorbox{colwp}{The hypothesis that} will be advanced here is that the biological counterpart of the CET value, in a given development stem cell, is represented by the ``configuration'' taken by a set of transposable elements active in somatic cells during development (like those responsible for the uneven pigmentation of the corns). We hypothesise that the TE configuration of a development stem cell is not fixed, but is modified in the course of development, so that different development stem cells have different TE configurations which, as CET values do in artificial development, represent the source of differentiation during natural development. As for the physical implementation of such configuration, there are broadly speaking two possibilities. While the CET is a single array, stored in a contiguous portion of the computer memory (see figure 7), transposons are located in many different positions of all chromosomes; the first possibility for the transposonic configuration is given by the set of all TE positions (figure 8). A second possibility is that the set of TE positions is the same for all development stem cells and that the configuration is implemented through the methylation pattern of such positions, different from cell to cell (figure 9). There is evidence to support both options, so that the most plausible mechanism is a combination of the two.

\begin{figure}[t!] \begin{center}
{\fboxrule=0.2mm\fboxsep=0mm\fbox{\includegraphics[width=8.00cm]{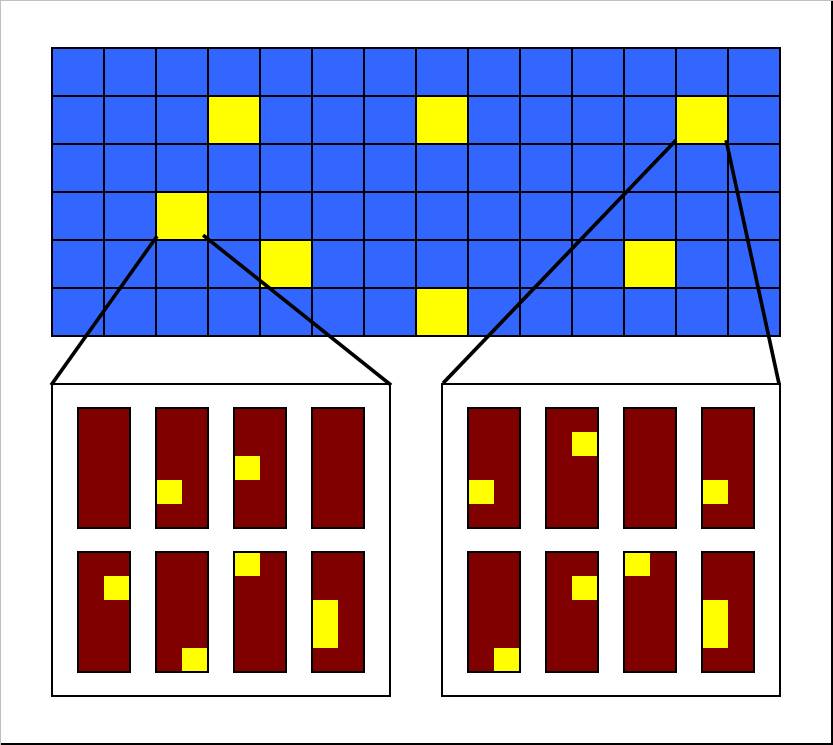}}}
\caption{Two different CET values in development stem cells, implemented by means of different patterns of TE positions on chromosomes.}
\label{figxx}
\end{center} \end{figure}

\begin{figure}[t!] \begin{center}
{\fboxrule=0.2mm\fboxsep=0mm\fbox{\includegraphics[width=8.00cm]{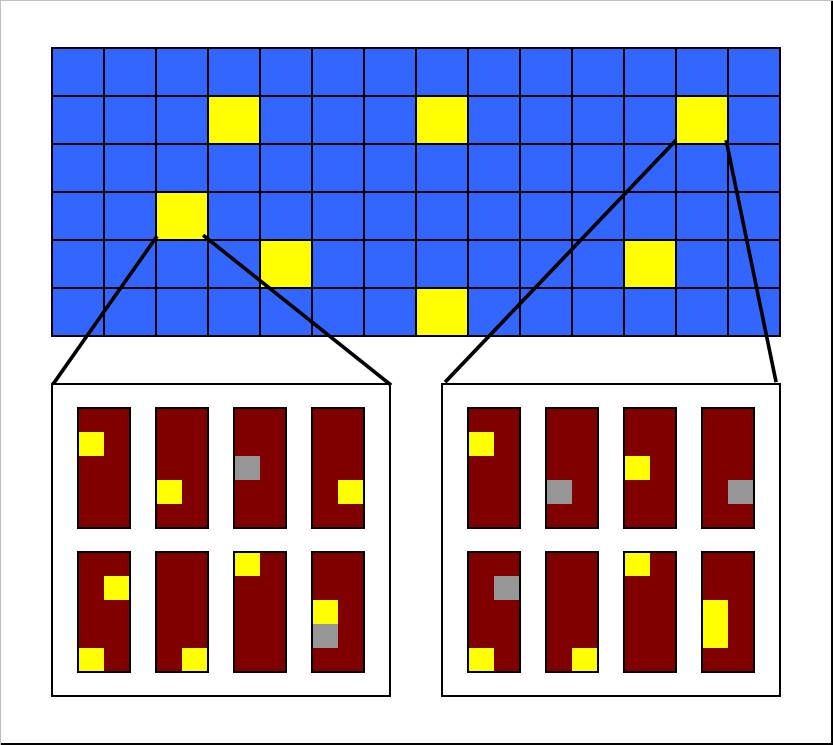}}}
\caption{Two different CET values in development stem cells, implemented by means of different methylation patterns superimposed on the same set of TE positions.}
\label{figxx}
\end{center} \end{figure}

\colorbox{colwp}{The interpretation provided} applies to the transposable elements active in somatic cells during development. As far as the transposons active in germline cells and with trans-generational effects, the proposed framework lead us to think of them as to the equivalent of the CET values copying themselves on instructions' XET values, in the biological counterpart of the Germline Penetration procedure. The name ``Germline Penetration'' draws inspiration from the imaginary path followed by the CET values generated during development which, metaphorically speaking, leave the driver cells they belong to and wander through the shape until they reach the equivalent of germline cells, which contain the genetic material that will be handed over to the subsequent generation; if the proposed interpretation is true, this picture is not to be considered as a useful metaphor, but as a faithful description of a key aspect of multicellular biology. In the light of this theory, the explanation of why sudden, burst-like waves of new classes of transposons spreading in genome is often associated to major evolutionary changes, becomes straightforward. In fact, as new structures are invented by evolution and become integrated in the animal's ``bauplan'', the corresponding set of new CET values generated spread into the Genome. Recalling how newly inserted instructions are set as inactive, it is not surprising to observe how the majority of transposable elements in genomes are rendered inactive through the mechanism of methylation.  

\colorbox{colwp}{If the theory} is correct, one key problem that needs to be addressed is by means of which biochemical mechanisms CET values are generated, i.e. which biochemical pathways make transposons jump in precise genomic locations during development. A possible scenario for the CET generation mechanism is the following. Each development stem cell emits in the neighborhood a mix of chemicals which represents the fingerprint of its own CET value; each normal cell keeps monitoring the concentrations of the said chemicals: if the strongest signal received is below a given threshold (meaning that the closest development stem cell is not close enough), the cell ``decides'' to turn itself into a development stem cell and the surrounding chemical mix, transduced into the nucleus, cause transposons to move around and insert in specific genome positions that, taken together, represent the CET value. As the new development stem cell starts diffusing its own fingerprint, nearby cells are automatically inhibited from undergoing the same transformation, so that the system reaches a state of equilibrium. 

\section{Conclusions}

\colorbox{colwp}{In the present} paper the evo-devo core of Epigenetic Tracking has been explored, and the result has been a novel hypothesis of the role of transposable elements. This hypothesis differentiates itself from previous ones by being grounded on a model of cellular growth which has been tested with computer simulations. The conclusion is that transposons have one of the most important roles in multicellular biology, as their distribution and methylation pattern in the genome of stem cells provides the key information necessary for cellular differentiation during development. At the end of development, they exit the cell and are trasported towards germline cells, where they become inserted into their genome. This mechanism represents the engine of evolutionary-developmental biology,  which lies at the heart of multicellar life. It is our belief that this framework will be of great value for the researchers who are trying to work out the biochemical details of transposable element biology.

\end{document}